\documentclass[12pt]{article}
\usepackage{amssymb}
\usepackage{epsfig}

\catcode`\@=11
\def\numberbysection{\@addtoreset{equation}{section}
        \def\theequation{\thesection.\arabic{equation}}}

\def\beq{\begin{equation}}
\def\eeq{\end{equation}}
\numberbysection
\begin{document}
\begin{titlepage}
\begin{center}
\hfill DFF  01/06/03 \\
\vskip 1.in {\Large \bf Gravity on a fuzzy sphere} \vskip 0.5in P.
Valtancoli
\\[.2in]
{\em Dipartimento di Fisica, Polo Scientifico Universit\'a di Firenze \\
and INFN, Sezione di Firenze (Italy)\\
Via G. Sansone 1, 50019 Sesto Fiorentino, Italy}
\end{center}
\vskip .5in
\begin{abstract}
We propose an action for gravity on a fuzzy sphere, based on a
matrix model. We find striking similarities with an analogous
model of two dimensional gravity on a noncommutative plane, i.e.
the solution space of both models is spanned by pure $U(2)$ gauge
transformations acting on the background solution of the matrix
model, and there exist deformations of the classical
diffeomorphisms which preserve the two-dimensional noncommutative
gravity actions.
\end{abstract}
\medskip
\end{titlepage}
\pagenumbering{arabic}
\section{Introduction}

A lot of knowledge has been successfully elaborated for
noncommutative gauge theories on a quantum hyper-plane, but the
ultimate goal remains to find consistent theories which unify
curved manifolds with noncommutative algebras \cite{1}-\cite{12}.
Usually theories of gravity are realized through the metric tensor
which can be thought as a field deforming the space-time from
flatness to a curved space. However one can write down gravity
theories whose basic fields deform the space-time starting from a
given curved manifold, for example in two dimensions a sphere.
This procedure will be applied in the present paper to the
noncommutative case.

Recently, in \cite{8} it has been suggested a model for
two-dimensional gravity on a noncommutative plane, made of gravity
fields and an auxiliary scalar field. However it is possible to
choose another noncommutative manifold as background, i.e. the
fuzzy sphere. The purpose of this article is to postulate the
action for noncommutative gravity on a fuzzy sphere and make
comparison with the results contained in \cite{8}, concerning the
flat case. We are aware of another work which has another proposal
on this subject \cite{12}, but we believe that our point of view
is fresh and more suited to the comparison with the flat case. We
start from a very simple matrix model action, based on $U(2)$
invariance, which contains the fuzzy sphere solution as a
background, and then construct the gravity fields as fluctuations
from this background. In our model the scalar field and gravity
fields are unified in three undifferentiated fluctuations. Only in
the commutative limit one can disentangle the gravity degrees of
freedom from the scalar one. We find that the solutions of this
model are made by pure $U(2)$ gauge transformations acting on the
background solution of the matrix model, and we explicitly study
some example. This property has a perfect analogy with the model
on the plane confirming that the two models of noncommutative
gravity are strictly related. Moreover we find that although the
classical diffeomorphism group is broken, it is possible to define
deformed diffeomorphisms which preserve the fuzzy sphere action,
in complete analogy with the quantum plane \cite{8}. Again
two-dimensional theories are good examples on which to test new
original ideas on noncommutative gravity.

\section{The model}

The fuzzy sphere \cite{13}-\cite{22} is a noncommutative manifold
represented by the following algebra:

\beq [ \hat{x}_i, \hat{x}_j ] = i \rho \epsilon_{ijk} \hat{x}_k \
\ \ \ \ \hat{x}^i = \rho L^i ,\label{21}\eeq

where $L^i$ is the usual angular momentum operator. The radius of
the sphere, obtained by the following condition

\beq \hat{x}_i \hat{x}_i = R^2 = \rho^2 L_i L_i = \rho^2 \frac{N (
N+2 )}{4} \label{22}\eeq

is kept fixed in the commutative limit $ N \rightarrow \infty $,
therefore

\beq \rho \sim \frac{1}{N} .\label{23}\eeq

We postulate that the action of gravity on a fuzzy sphere is given
by the following action :

\beq S = \frac{1}{g^2} Tr [ \frac{i}{3} \epsilon^{ijk} X_i X_j X_k
+ \frac{\rho}{2} X_i X_i ] ,\label{24}\eeq

where $X_i$ is not only an $(N+1) \times (N+1)$ hermitian matrix,
but it has also values in $U(2)$:

\beq X_i = X_i^A t_A = X^0_i + X^a_i \tau_a \label{25}\eeq

where $\tau_a$ are the Pauli matrices. Having chosen the
generators of the $U(2)$ algebra hermitian , all the component
matrices $X^0_i$, $X^a_i$ are therefore $(N+1) \times (N+1)$
hermitian matrices.

Let us note the analogy of this action with the proposed action
for gravity on a noncommutative two-dimensional plane \cite{8}

\beq S_p = Tr [ \phi \epsilon^{ij} ( [ X_i, X_j ] - i
\theta_{ij}^{-1} ) ] .\label{26}\eeq

In this case there are also three independent hermitian operators
as in our case, but we note that a pure scalar operator $\phi$ is
disentangled from the true gravity degrees of freedom, the two
matrices $X_i ( i=1,2 )$. In our case (\ref{24}) we can think that
the scalar operator $\phi$ and the gravity degrees of freedom are
unified in our operators $X_i ( i=1,2,3 )$ and that it is possible
to disentangle the scalar field from the gravity fields only in
the commutative limit $\rho \rightarrow 0$ (\ref{23}).

Let us describe in details how to obtain such a picture. Firstly
we need to develop the matrix $X_i$ as a background plus
fluctuations ( see also \cite{14}-\cite{17} ):

\beq X_i = \hat{x}_i + \rho R \hat{a}_i .\label{27}\eeq

In front of the $U(2)$ noncommutative connection $\hat{a}_i$ there
is a factor $\rho$, which means that in the $N \rightarrow \infty$
limit the difference between $X_i$ and the background $\hat{x}_i$
must be negligible. The action $S$ (\ref{24}) is invariant under
the $U(2)$ unitary transformation

\beq X_i \rightarrow U^{-1} X_i U \label{28}\eeq

which extends the local Lorentz invariance of the gravity theory
to a global symmetry of the matrix model.

By developing $U$ in terms of an infinitesimal transformation

\beq U \sim 1 + i \hat{\lambda} \label{29}\eeq

the fluctuations around the fixed background transform as

\beq \hat{a}_i \rightarrow \hat{a}_i - \frac{i}{R} [ \hat{L}_i,
\hat{\lambda} ] + i [ \hat{\lambda}, a_i ] .\label{210}\eeq

To reformulate the action in terms of the fluctuations $\hat{a}_i$
we have to define a star product on the fuzzy sphere, analogous to
the Moyal star product for the plane.

Recall that a matrix on the fuzzy sphere can be developed in terms
of the noncommutative analogue of the spherical harmonics
$\hat{Y}_{lm}$:

\beq \hat{Y}_{lm} = R^{-l} \sum_a f^{(lm)}_{a_1, a_2,...a_l} \
\hat{x}_{a_1} ... \hat{x}_{a_l} \label{211}\eeq

while the classical spherical harmonics are defined with
$\hat{x}_i$ substituted with the commutative coordinates $x_i$.

A general hermitian matrix

\beq \hat{a} = \sum^{N}_{l=0} \sum^{l}_{m=-l} \ a_{lm} \
\hat{Y}_{lm} \ \ \ \ \ \ a^{*}_{lm} = a_{l -m} \label{212}\eeq

corresponds therefore to an ordinary function on the commutative
sphere as:

\beq a( \Omega ) = \frac{1}{N+1} \sum^{N}_{l=0} \sum^{l}_{m=-l} Tr
( \hat{Y}_{lm}^{\dagger} \hat{a} ) Y_{lm} ( \Omega )
\label{213}\eeq

and the ordinary product of matrices is mapped to the star product
on the commutative sphere:

\begin{eqnarray}
& \ & \hat{a} \hat{b}  \rightarrow  a * b \nonumber \\
& \ & a(\Omega) * b(\Omega) = \frac{1}{N+1} \sum^{N}_{l=0}
\sum^{l}_{m=-l} Tr ( \hat{Y}^{\dagger}_{lm} \hat{a} \hat{b} )
Y_{lm} ( \Omega ) . \label{214}
\end{eqnarray}

Derivative operators can be constructed using the adjoint action
of $\hat{L}_i$ and tend to the classical Lie derivative $L_i$ in
the $N \rightarrow \infty$ limit:

\beq Ad( \hat{L}_i ) \rightarrow L_i = \frac{1}{i} \epsilon_{ijk}
x_j \partial_k . \label{215}\eeq

$L_i$ can be expanded in terms of the Killing vectors of the
sphere

\beq L_i = - i K_i^a \partial_a . \label{216}\eeq

In terms of $K_i^a$ we can form the metric tensor $g_{ab} = K^i_a
K^i_b$. The explicit form of these Killing vectors is

\begin{eqnarray}
K_1^{\theta} & = & - sin \phi \ \ \ \ \  K^{\phi}_1 = - cotg
\theta cos \phi \nonumber \\
K_2^{\theta} & = &  cos \phi \ \ \ \ \  K^{\phi}_1 = - cotg
\theta sin \phi \nonumber \\
K_3^{\theta} & = & 0  \ \ \ \ \ \ \ K_3^{\phi} = 1 \label{217} .
\end{eqnarray}

Trace over matrices can be mapped to the integration over
functions:

\beq \frac{1}{N+1} Tr ( \hat{a} ) \rightarrow \int
\frac{d\Omega}{4\pi} \ a( \Omega ) . \label{218}\eeq

Having introduced the star product (\ref{214}), we can map the
action $S(\lambda)$ to the following field theory action as in
\cite{14}-\cite{17}

\begin{eqnarray}
S & = & \frac{i}{4 g^2_f} \epsilon_{ijk} Tr \int d\Omega ( (L_i
a_j) a_k + \frac{R}{3} [a_i,a_j] a_k - \frac{i}{2} \epsilon_{ijl}
a_l a_k )_{*} \nonumber \\
& + & \frac{\pi}{12 g^2_f} \frac{N(N+2)}{R^2}
\label{219}\end{eqnarray}

where the residual trace is for the $U(2)$ degrees of freedom. The
classical limit is realized as

\beq R = {\rm fixed} \ \ \ \ g^2_f = \frac{ 4 \pi g^2 }{(N+1)
\rho^4 R^2} = {\rm fixed} \ \ \ \ N \rightarrow \infty .
\label{220}\eeq

In the commutative limit, the star product becomes the commutative
product, and the scalar field $\phi$ is separable from the gravity
fields as in the following formula

\beq R a_i (\Omega) = K^a_i A_a ( \Omega) + \frac{x_i}{R} \phi(
\Omega ) \label{221}\eeq

where $A_a (\Omega)$ are the gravity fields on the sphere.

Defining the gauge covariant field strength $\hat{F}_{ij}$ as

\begin{eqnarray}
\hat{F}_{ij} & = & \frac{1}{\rho^2 R^2} ( [X_i, X_j] - i \rho
\epsilon_{ijk} X_k ) \nonumber \\
& = & [ \frac{\hat{L}_i}{R}, \hat{a}_j ] - [ \frac{\hat{L}_j}{R},
\hat{a}_i ] + [ \hat{a}_i, \hat{a}_j ] - \frac{i}{R}
\epsilon_{ijk} \hat{a}_k \label{222}\end{eqnarray}

it can be expanded, only in the commutative limit  in terms of the
components fields $(\phi, A_a)$ (\ref{221}) as follows:

\beq F_{ij} (\Omega) = \frac{1}{R^2} K_i^a K_j^b F_{ab} +
\frac{i}{R^2} \epsilon_{ijk} x_k \phi + \frac{1}{R^2} x_j K_i^a
D_a \phi - \frac{1}{R^2} x_i K_j^a D_a \phi \label{223}\eeq

where $F_{ab} = - i ( \partial_a A_b - \partial_b A_a ) + [A_a,
A_b] $ and $D_a = - \partial_a + [A_a, ... ] $ for the $U(2)$
case.

The classical limit of the action $S$ is determined to be:

\begin{eqnarray}
S & = & \frac{1}{2g^2_f R^2} Tr \int d\Omega ( i \epsilon_{ijk}
K_i^a K_j^b F_{ab} \frac{x_k}{R} \phi - \phi^2 ) = \nonumber \\
& = & \frac{1}{2 g^2_f R^2} Tr \int d\Omega (
\frac{i\epsilon^{ab}}{\sqrt{g}} F_{ab} \phi - \phi^2 ) \label{224}
\end{eqnarray}

where $\epsilon^{ab}$ is defined as $\epsilon^{\theta\phi} =1$.

The corresponding equations of motion are the null condition of
the field strength

\beq \hat{F}_{ij} \propto [ X_i, X_j ] - i \rho \epsilon_{ijk} X_k
= 0 . \label{225} \eeq

In the classical limit these correspond to

\begin{eqnarray}
& \ & i \epsilon^{ab} F_{ab} - 2 \sqrt{g} \phi = 0 \nonumber \\
& \ & D_a \phi = 0 . \label{226}
\end{eqnarray}

Let us discuss them in components. Firstly we identify the gravity
fields $A_a$ as:

\beq A_a = e_a^1 \tau_1 + e_a^2 \tau_2 + \omega^3_a \tau_3 +
\omega_a^0 \label{227}\eeq

where $e_a^i (i=1,2)$ are the two components of the vierbein,
$\omega_a^3$ is the usual abelian two-dimensional spin connection
and $\omega_a^0$ is an extra scalar spin connection which appears
for consistency in the noncommutative case.

To compute the classical equations of motion (\ref{226}) in
components we need:

\beq F_{ab} = F_{ab}^A \tau_A = F_{ab}^1 \tau_1 + F_{ab}^2 \tau_2
+ F_{ab}^3 \tau_3 + F_{ab}^0 \label{228}\eeq

where

\begin{eqnarray}
F_{ab}^{\alpha} & = & -i ( \partial_a e^{\alpha}_b - \partial_b
e^{\alpha}_a ) + i \epsilon^{\alpha\beta} ( \omega^3_a e^{\beta}_b
- \omega^3_b e^{\beta}_a ) \ \ \ \ \alpha = 1,2 \nonumber \\
F_{ab}^3 & = & \partial_a \omega^3_b - \partial_b \omega^3_a + i
\epsilon^{\alpha\beta} e^{\alpha}_a e^{\beta}_b \nonumber \\
F_{ab}^0 & = & \partial_a \omega_b^0 - \partial_b \omega_a^0
\label{229}\end{eqnarray}

Analogously the scalar field $\phi$ has an expansion

\beq \phi = \phi^A \tau_A = \phi^0 + \phi^a \tau_a \ \ \ \ a=
1,2,3 . \label{230}\eeq

Finally we obtain :

\begin{eqnarray}
& \ & i \epsilon^{ab} F_{ab}^A - 2 \sqrt{g} \phi^A = 0 \nonumber \\
& \ & D_a \phi = 0 . \label{231}
\end{eqnarray}

\section{Classical solutions of the model}

Let us discuss the classical solutions of the model. We will find
in this respect a striking similarity with the corresponding model
on the plane. Recalling that the model (\ref{26}) has as equations
of motions the following ones:

\begin{eqnarray}
& \ & [ X_i, \phi ] = 0 \nonumber \\
& \ & [ X_i, X_j ] = i \theta_{ij}^{-1} . \label{31}
\end{eqnarray}

A particular solution to this equation, which we call background
solution is given by the choice

\begin{eqnarray}
X_i & = & \hat{p}_i \ \ \ \  \ \hat{p}_i = - i \theta^{-1}_{ij}
\hat{x}_j \nonumber \\
\phi_0 & = & \phi_0^0 + \phi^a_0 \tau_a = \phi^A_0 \tau^A \ \ \ \
[\hat{p}_i , \phi^A_0 ] = 0 \label{32}
\end{eqnarray}

with the scalar components all constants ( $\partial_i \phi^A_0 =
0 $ ).

The general solution of this model can be found by applying a pure
$U(2)$ gauge transformation on the background solution (\ref{32})
as follows

\begin{eqnarray}
\phi & = & U^{-1} \phi_0 U \nonumber \\
X_i & = & U^{-1} \hat{p}_i U \label{33}
\end{eqnarray}

where $U$ is an unitary operator, with values in the group $U(2)$.

The corresponding equations of motions for the model on the fuzzy
sphere are again solved by a pure gauge transformation acting on
the background of the matrix model

\begin{eqnarray}
& \ & \hat{F}_{ij} \propto [ X_i, X_j ] - i \rho \epsilon_{ijk}
X_k = 0
\nonumber \\
& \ & X_i = U^{-1} \hat{x}_i U . \label{34}
\end{eqnarray}

As an exercise, let us compute an example of such solutions, and
identify it with a classical solution of the corresponding gravity
theory on a sphere. Let us pose

\beq X_i = \hat{x}_i + ( U^{-1} \hat{x}_i U - \hat{x}_i ) =
\hat{x}_i + A_i . \label{35}\eeq

The field $A_i$ must be a fluctuation of order $\rho$, as
discussed at the beginning. Let us compute the solution generated
by the following unitary operator for the simplest case $N=2$,
where $\hat{x}_i$ is proportional to the Pauli matrices:

\beq U = e^{2i\alpha n^1_i \hat{x}_i} P + e^{2i\beta n^2_i
\hat{x}_i} ( 1- P ) = U_L P + U_R ( 1- P ) \label{36}\eeq

where $\alpha, \beta$ are pure constants, $n^1_i, n^2_i$ are two
generic unit vectors, $\hat{x}_i$ are proportional to the Pauli
matrices and the group factor is a projector

\beq P = \frac{ 1+ n_i \tau_i }{2} \label{37}\eeq

with $n_i$ another unit vector.

A simple algebra shows that

\begin{eqnarray}
A_i & = & [ - 2 sin (\rho \alpha ) cos ( \rho \alpha )
\epsilon_{ijk} n^1_j \hat{x}_k + 2 sin^2 ( \rho \alpha ) ( n^1_i (
n^1_j \hat{x}_j ) - \hat{x}_i ) ] P + \nonumber \\
& + & [ - 2 sin (\rho \beta ) cos ( \rho \beta ) \epsilon_{ijk}
n^2_j \hat{x}_k + 2 sin^2 ( \rho \beta ) ( n^2_i ( n^2_j \hat{x}_j
) - \hat{x}_i ) ] ( 1- P ) . \label{38}
\end{eqnarray}

It is possible to check directly that $A_i$ satisfies the
equations of motion

\beq [ \hat{x}_i, A_j ] - [ \hat{x}_j, A_i ] + [ A_i, A_j ] = i
\rho \epsilon_{ijk} A_k . \label{39}\eeq

By developing the matrix $A_i$ for $N\rightarrow \infty$ ( $\rho
\rightarrow 0 $ ), it is easy to show that $A_i \sim O(\rho)$  as
in (\ref{27}) , confirming that it is possible to generate,
through unitary transformations, fluctuations which tend to a
nontrivial classical limit.

In the commutative limit the fluctuation must be truncated to
order $O(\rho)$:

\beq A_i = - 2 \rho \alpha \epsilon_{ijk} n^1_j x_k P - 2 \rho
\beta \epsilon_{ijk} n^2_j x_k ( 1 - P ) \label{310}\eeq

and it satisfies the classical equation

\beq L_i A_j - L_j A_i = i \epsilon_{ijk} A_k \ \ \ \ \ L_i = -i
\epsilon_{ijk} x_j \frac{\partial}{\partial x_k} . \label{311}\eeq

In fact the action of the first member gives rise to

\begin{eqnarray} L_i A_j - L_j A_i & = & - 2i \rho \alpha  ( n^1_i x_j - n^1_j
x_i ) P - 2i \rho \beta  ( n^2_i x_j - n^2_j x_i ) ( 1- P )=
\nonumber \\ & = & i \epsilon_{ijk} A_k . \label{312}
\end{eqnarray}

This truncated fluctuation (\ref{310}) satisfies the gauge
condition

\beq x_i A_i = 0 \label{313}\eeq

and therefore it will not contribute to the scalar field, but only
to the gravity fields. By developing

\beq A_i = \frac{x_i}{R} \phi + K_i^a A_a \label{314}\eeq

we can extract the gravity fields as

\beq A_a = g_{ab} K_i^b A_i = \sqrt{g} \epsilon_{ab} [ ( - 2 \rho
\alpha n^1_i K_i^b ) P + ( - 2 \rho \beta n^2_i K_i^b ) ( 1- P )].
\label{315}\eeq

The classical equations of motion

\begin{eqnarray}
F_{ab} & = & \partial_a A_b - \partial_b A_a + [ A_a, A_b ] = 0
\nonumber \\
\phi & = &  0 \label{316}
\end{eqnarray}

are solved since $[A_a, A_b]=0$, thanks to $[P,P] = [ P, 1-P ] = 0
$ and since

\beq \partial_\theta A_\phi = \partial_\phi A_\theta = sin \theta
[ 2 \rho \alpha ( n^1_1 cos \phi + n^1_2 sin \phi ) + 2 \rho \beta
 ( n^2_1 cos \phi + n^2_2 sin \phi )] . \label{317}\eeq

 \section{Deformed diffeomorphisms on the fuzzy sphere}

 Let us recall the steps that are necessary to prove the existence
 of deformed diffeomorphisms on the noncommutative plane. In the
 commutative limit, the action for gravity on a plane is invariant
 not only under local Lorentz invariance, but also under
 infinitesimal diffeomorphisms ( Lie derivatives ) along an
 arbitrary vector field $ v = v^{\mu} \partial_\mu $

\begin{eqnarray}
{\cal L}_v A_i & = & ( d i_v + i_v d ) A_i \ \ \ \  A_i = X_i - \hat{p}_i \nonumber \\
{\cal L}_v \phi &  = & i_v d \phi \label{41}
\end{eqnarray}

where $i_v$ is the inner product on differential forms. Using the
Leibnitz rule for the inner product is is possible to prove that

\begin{eqnarray}
{\cal L}_v A & = & \delta_{i_v A} A  + i_v F \nonumber \\
{\cal L}_v \phi & = & \delta_{i_v A} \phi + i_v D \phi \label{42}
\end{eqnarray}

the infinitesimal diffeomorphisms contains a gauge transformation
with parameter $\lambda = i_v A$, and a residual part which is
vanishing on shell ( i.e. on the equations of motion ).

In the noncommutative case the situation is quite different. It is
easy to generalize the local Lorentz invariance, but it seems
quite difficult to define the corresponding infinitesimal
diffeomorphisms. However in the paper \cite{8} this question has
been solved for the noncommutative plane. We are going to prove
that the same property holds also for our action (\ref{24}) on the
fuzzy sphere, but we recall the solution on a noncommutative plane
as a suggestion.

Let us generalize the inner product in the noncommutative case;
starting from a $p$-form $\omega^p$ we define

\begin{eqnarray}
& \ & \omega^p = \frac{1}{p!} \omega_{\mu_1...\mu_p} dx^{\mu_1}
\wedge .... \wedge dx^{\mu_p} \nonumber \\
& \ & i^{*}_v \omega^p = \frac{1}{2 (p-1)!} [ v^\rho *
\omega^p_{\rho \mu_1...\mu_{p-1}} + \omega^p_{\rho
\mu_1...\mu_{p-1}} * v^\rho ] dx^{\mu_1} \wedge .... \wedge
dx^{\mu_p} . \label{43}
\end{eqnarray}

In this case the Leibnitz rule is not valid anymore for $i_v^{*}$.
We then define the deformed diffeomorphisms on the noncommutative
plane as

\begin{eqnarray}
& \ &  \Delta^{*}_v = i^{*}_v D + \delta_{i_v^{*} A} \nonumber \\
& \ &  \Delta^{*}_v A = i^{*}_v F + \delta_{i_v^{*} A} A \nonumber \\
& \ &  \Delta^{*}_v \phi = i^{*}_v D \phi + \delta_{i_v^{*} A}.
\label{44}
\end{eqnarray}

Since it is known that the action is invariant under local Lorentz
transformations, in order to prove its invariance under the
transformation (\ref{44}) it is enough to prove it for the
residual transformation:

\begin{eqnarray}
& \ & \delta^{'}_v A = i_v^{*} F \nonumber \\
& \ & \delta^{'}_v \phi = i_v^{*} D \phi . \label{45}
\end{eqnarray}

Let us reformulate it in terms of the matrix model variables
$(\phi, X_i (i=1,2))$:

\begin{eqnarray}
& \ & S = Tr ( \phi \epsilon^{ij} F_{ij} ) \ \ \ \ \ F_{ij} = [
X_i,
X_j ] - i \theta_{ij}^{-1} \nonumber \\
& \ & \delta^{'}_v X_i = v^\alpha F_{i\alpha} + F_{i\alpha}
v^\alpha \nonumber \\
& \ & \delta^{'}_v \phi = v^\alpha [ X_\alpha, \phi ] + [
X_\alpha, \phi ] v^\alpha . \label{46}
\end{eqnarray}

Then the variation of the action is given by

\begin{eqnarray}
\delta^{'} S & = & Tr( \delta \phi \epsilon^{ij} F_{ij} + 2
\epsilon^{ij} \delta X_i [ X_j, \phi ] ) = \nonumber \\
& = & Tr (( \epsilon^{ij} v^\alpha + \epsilon^{j\alpha} v^i +
\epsilon^{\alpha i} v^j ) ( [ X_\alpha, \phi ] F_{ij} + F_{ij} [
X_\alpha, \phi ] )) .\label{47}
\end{eqnarray}

The tensor under parenthesis is zero, due to the cyclic dependence
from three indices which can take only two values. Therefore it is
proved that (\ref{45}) is indeed a deformed diffeomorphism on the
noncommutative plane.

Now comes the interesting part since we have found for our
postulated action (\ref{24}) an analogous property, i.e. it is
possible to generalize the diffeomorphism group of classical
gravity on the sphere to a nontrivial deformed diffeomorphism
group on a fuzzy sphere.

We start directly from the matrix variable $X_i (i=1,2,3)$ of our
model

\beq S = \frac{1}{g^2} Tr ( \frac{i}{3} \epsilon^{ijk} X_i X_j X_k
+ \frac{\rho}{2} X_i X_i ). \label{48}\eeq

We note that the variation of $S$ is vanishing on shell

\beq \delta S \propto \frac{i}{g^2} Tr [ \epsilon^{ijk} ( \partial
X_i \hat{F}_{jk} ) ] .\label{49}\eeq

We define the deformed diffeomorphism as generated by

\begin{eqnarray}
& \ & \delta^{'}_{v} X_i = v^{\alpha} \hat{F}_{i \alpha} +
\hat{F}_{i \alpha}
v^{\alpha} \nonumber \\
& \ & \hat{F}_{ij} \propto [X_i, X_j ] - i \rho \epsilon_{ijk} X_k
. \label{410}
\end{eqnarray}

Let us note the nontrivial dependence on $\rho $ of the deformed
rules which tends to zero in the commutative limit. Therefore we
obtain for the generic variation (\ref{49}) the following result

\beq \delta^{'}_v S = \frac{i}{g^2} Tr [ v^\alpha \epsilon^{ijk} (
\hat{F}_{i \alpha} \hat{F}_{jk} + \hat{F}_{jk} \hat{F}_{i \alpha}
) ] .\label{411}\eeq

This particular variation can be rewritten as

\begin{eqnarray}
& \ & \delta^{'}_v S = \frac{i}{g^2} Tr [ ( v^\alpha
\epsilon^{ijk} - v^i \epsilon^{jk\alpha} - v^j \epsilon^{ki\alpha}
- v^k
\epsilon^{ij\alpha} ) \nonumber \\
& \ & ( \hat{F}_{i \alpha} \hat{F}_{jk} + \hat{F}_{jk}
\hat{F}_{i\alpha} ) ] \label{412}
\end{eqnarray}

and again the variation is proportional to a tensor which is zero
because of the cyclic dependence from more indices than the number
of values. It is this property which finally permits to extend the
diffeomorphism group for all the known two-dimensional
noncommutative manifolds.

We note that to reach this result the choice of the action is
crucial and that this property doesn't hold for the general matrix
model action of the fuzzy sphere, defined in \cite{16}-\cite{17}.

\section{Conclusion}

In this article we have introduced in two dimensions an
alternative description of noncommutative gravity starting from
the fuzzy sphere background, instead of the noncommutative plane.
We have found several properties in common between the two models
which may suggest their equivalence.

Let us recapitulate them; the classical solutions of the two
models are made by pure $U(2)$ unitary transformations acting on
the background solutions, and the diffeomorphism group can be
deformed in both cases to a nontrivial noncommutative invariance,
being the first examples of general covariance in a matrix model.

We can think the solution space of noncommutative gravity as an
extension of the simplest noncommutative algebras ( quantum plane
and fuzzy sphere ) to a generic two-dimensional noncommutative
manifold. Indeed the fuzzy sphere solution can be found by
deforming the quantum plane with non trivial gravity fields
\cite{8}.

A possible generalization of the present research would be to
extend these results to the case of four dimensions
\cite{23}-\cite{24}, in which however the issue of a deformed
general covariance is a harder problem.

Returning to two dimensions, we would like to point out that our
postulate action is probably unique if we demand the presence of
deformed diffeomorphisms, and previous proposals on this subject
\cite{12} are surely different because of the lack of this
property. Moreover in our model the scalar field and the gravity
fields are unified in three undifferentiated fluctuations,
resulting in a more symmetric description. Only in the commutative
limit one recovers the distinction between the scalar field and
the gravity fields. The fuzzy sphere case is also interesting
because it gives a finite description of noncommutative gravity (
with a function space truncated to a finite Hilbert space ).
Therefore it would be interesting to study the quantum properties
of this model, as a noncommutative counterpart of two dimensional
quantum gravity which has been deeply studied.

\end{document}